\documentclass{article}

\usepackage{PRIMEarxiv}

\usepackage[utf8]{inputenc} 
\usepackage[T1]{fontenc}    
\usepackage{hyperref}       
\usepackage{url}            
\usepackage{booktabs}       
\usepackage{amsfonts}       
\usepackage{nicefrac}       
\usepackage{microtype}      
\usepackage{fancyhdr}       
\usepackage{graphicx}       
\usepackage{natbib}         
\usepackage{listings}       
\usepackage{xcolor}         
\usepackage{caption}        
\usepackage{float}          
\graphicspath{{assets/img/2026-04-27-coding-agents/}} 

\definecolor{codebg}{rgb}{0.95,0.95,0.95}
\definecolor{codegreen}{rgb}{0.25,0.50,0.35}
\definecolor{codegray}{rgb}{0.5,0.5,0.5}
\definecolor{codepurple}{rgb}{0.58,0,0.82}

\title{Can Coding Agents be General Agents?}

\author{
  Maksim Ivanov\thanks{Corresponding author: \texttt{maksim@agenticlabs.com}} \\
  Agentic Labs \\
  \And
  Abhijay Rana \\
  Agentic Labs \\
  \And
  Gokul Prabhakaran \\
  Agentic Labs \\
}

\begin{document}
\raggedbottom
\maketitle

\begin{abstract}
As coding agents have seen rapid capability and adoption gains, users are applying them to general tasks beyond software engineering. In this post, we investigate whether coding agents can successfully generalize to end-to-end business process automation. We identify gaps in current evaluations, and conduct a case study to evaluate a coding agent on practical business tasks in an open-core Enterprise Resource Planning system. We find that the agent reliably completes simple tasks but exhibits characteristic failures on complex tasks, suggesting that bridging domain logic and code execution is a key bottleneck to generalizability.
\end{abstract}

\section{Introduction}

Immediately after the emergence of Transformer-based Language Models (LMs), researchers and developers began exploring LM code generation \cite{ahmad2021plbart}. With heavy investment into training models specifically on coding data, LMs have seen dramatic improvements in coding capabilities \cite{chen2021codex}. The top score on SWE-Bench Verified, a benchmark testing models on real-world software engineering tasks, jumped from 49\% to 78\% through 2025 \cite{jimenez2023swebench, openai2024swebenchverified}. The newer Terminal Bench, released in May of 2025, has also tracked frontier LLMs improving from 43\% to 61\% on complex tasks in the terminal, including analyzing data, calling APIs, and addressing security vulnerabilities \cite{terminalbench2024}.

Propelled by compounding improvement, frontier labs have been developing coding agents, which augment foundation models with a shell sandbox and code editor to help with coding tasks. These agents, including Claude Code, Codex CLI, and Gemini CLI, have seen explosive developer adoption: since launching in May 2025, Claude Code now receives over 4.4 million downloads every week \cite{anthropic2025claudecode}.

Surprisingly, people are using these coding agents for purposes far beyond the realm of software development. Users report applying coding agents to tax preparation, creating content, personal knowledge management, and more \cite{steipete2025claudecomputer}. At their core, ``coding'' agents are versatile: even Anthropic has acknowledged this shift, rebranding its Claude Code SDK to the general ``Agent SDK'' \cite{anthropic2024agentsdk}. This makes sense: knowledge work takes place entirely in software---business apps, browsers, spreadsheets, databases. Coding agents are naturally fit to meet knowledge work where it lives.

\subsection{How Does a Coding Agent Work?}

Broadly, an ``AI Agent'' is a system that autonomously pursues a goal by perceiving its environment, reasoning iteratively, and taking actions, with minimal human intervention. Typically, agents take action through manually predefined tools \cite{schick2023toolformer}. Often, these tools include complex business logic that reduces the reasoning burden on the LM itself. For example, a simple banking agent may be provided: ``read\_balance,'' ``withdraw\_money,'' ``deposit\_money,'' and ``transfer\_funds.'' In practice, each of these tools would include guardrails to prevent prohibited transactions or incorrect calculations.

Coding agents are a specific type of agent designed for software engineering tasks. These agents operate within software development environments (repositories, IDEs, sandboxes, terminals). \emph{Coding agents are unique because they write, execute, and debug their own scripts at runtime, instead of being limited to a pre-defined set of tools.} They are especially versatile: they can quickly orient themselves in new software environments by querying for information, installing packages to unlock new capabilities, and resolving errors from logs. This self-governed feedback loop unlocks utility beyond software development: coding agents can theoretically plug-and-play into \emph{any} software environment---offering an interesting pathway to generalizability.

\begin{figure}[ht]
    \centering
    \includegraphics[width=0.4\textwidth]{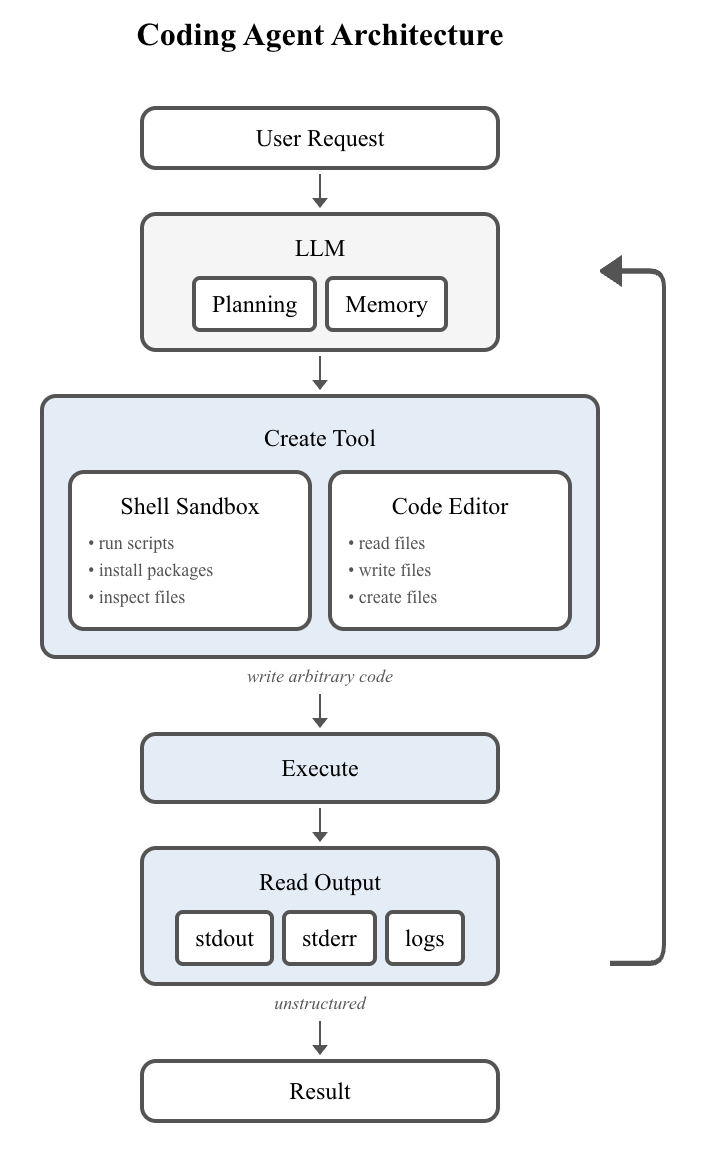}
    \caption{The coding agent is an emergent agent architecture, which allows a model to create its own code tools at runtime and iterate with the help of code outputs and error logs.}
    \label{fig:architecture}
\end{figure}

Experimentally, past research has shown that training models on code improves general reasoning. For instance, Aryabumi et al.\ find that adding code data to pre-training yields up to 8.2\% relative improvement on natural language reasoning benchmarks \cite{aryabumi2024codepretraining}. Separately, Wang et al.\ demonstrate that their CodeAct agent, using executable Python scripts, outperformed JSON and text-based tool-calling alternatives with a 20\% higher success rate on non-coding tasks \cite{wang2024codeact}. Furthermore, the continued scaling of test-time compute means coding agents are increasingly adept at attacking multi-hour, project-scale tasks \cite{openai2025gpt51}. Practically and experimentally, there is visible potential in applying coding agents to general tasks.

Given this convergence, we ask: \textbf{Can coding agents succeed as general business agents? And if not yet, where do they break down?}

This post makes three contributions:

\begin{enumerate}
    \item \textbf{A framework for coding-agent generalization.} We propose that success as a general business agent requires bidirectional translation between business/domain and code/software layers, and articulate four concrete capabilities this entails.
    \item \textbf{An evaluation gap analysis.} We survey prominent benchmarks and show that code-level evals (SWE-bench, Terminal-Bench) lack business context while domain-reasoning evals ($\tau$-bench \cite{yao2024taubench}, BFCL \cite{gorilla2024bfcl}) lack complex code execution---leaving full-stack translation underexplored.
    \item \textbf{Observations from an ERP case study.} We deploy frontier coding agents on a production-grade Enterprise Resource Planning (ERP) software instance with multi-constraint Sales-to-Fulfilment and HR operational tasks, document distinct failure modes at the business-code boundary, and propose asymmetric feedback from the environment to explain persistent agent overconfidence.
\end{enumerate}

\section{Defining Success}

\textbf{In order to be a reliable general agent, a coding agent must be excellent at translating between the business/domain layer and the code/software layer.} In practice, that means it can:

\begin{enumerate}
    \item Understand business-level requests and policies: turning instructions like ``approve this expense if it fits our travel policy and budget'' into precise goals and constraints.
    \item Inspect the current system state via generated and executed code: writing and running queries or scripts to see what requests, data, and approvals already exist across the relevant systems.
    \item Plan a business-level solution: deciding, at the domain level, what should actually happen (approve, modify, escalate, or reject) given the policies and current state.
    \item Encode that plan back into the software: implementing the decision as code/API calls that update the system state to match the intended business outcome.
\end{enumerate}

\section{What Gets Measured Gets Improved}

It's worth examining how we evaluate frontier models and agents today. \textbf{Current evaluations largely fall into two camps: those that test code-level competence and those that test business/domain-level reasoning.} The full-stack, bi-directional \emph{translation} between them is undercovered.

\subsection{Code and System-Level Benchmarks}

\textbf{SWE-Bench} tests whether models can resolve real GitHub issues by generating code patches. Given a repository snapshot and an issue description like \emph{``DatetimeIndex.to\_period() fails with timezone-aware timestamps''}, the agent must locate the bug across thousands of files, generate a fix, and pass the repository's unit tests. The input is technical; the output is technical; the evaluation is whether tests pass. SWE-Bench has become a standard because it captures real software engineering difficulty, but the ``business context'' facet is thin.

\textbf{Terminal-Bench} takes a similar approach for evaluating terminal mastery. Tasks range from \emph{``Build Linux kernel 6.9 from source with a custom printk message''} to \emph{``Configure a git server that pushes to a webserver on port 8080.''} Agents interact with a sandboxed Linux terminal and are evaluated with task-specific checks: whether files exist, services respond, and commands succeed. Again, the request is system-level, and the evaluation is system-level.

\subsection{Tool-Use and Domain Reasoning Benchmarks}

Tool-use benchmarks like \textbf{The Berkeley Function Calling Leaderboard (BFCL)} evaluate whether a model can turn prompts and pre-defined function specifications into \emph{correct function calls} across many domains. For example, if tasked with calculating a five-year compounding return and given a tool set containing a ``calculate\_compound\_interest'' function, the agent must locate that tool, identify the principal, interest rate, and time frame from context, and correctly pass them as arguments into the function.

BFCL measures how often models get tool names and arguments right, probing the flow from business requests to tool selection and arguments, but without complex code execution or policy adherence behind it.

\textbf{$\tau$-bench} pushes further by adding policy compliance. An airline agent may handle a user request: \emph{``I want to change my reservation to a different destination.''} The agent must gather information through dialogue, consult policy documents about change fees and cabin-class restrictions, and call domain APIs correctly to make the change. This is close to what a real deployment would require, but the underlying, synthetically-generated database and API are intentionally much simpler than real-world business software.

\begin{figure}[ht]
    \centering
    \includegraphics[width=0.65\textwidth]{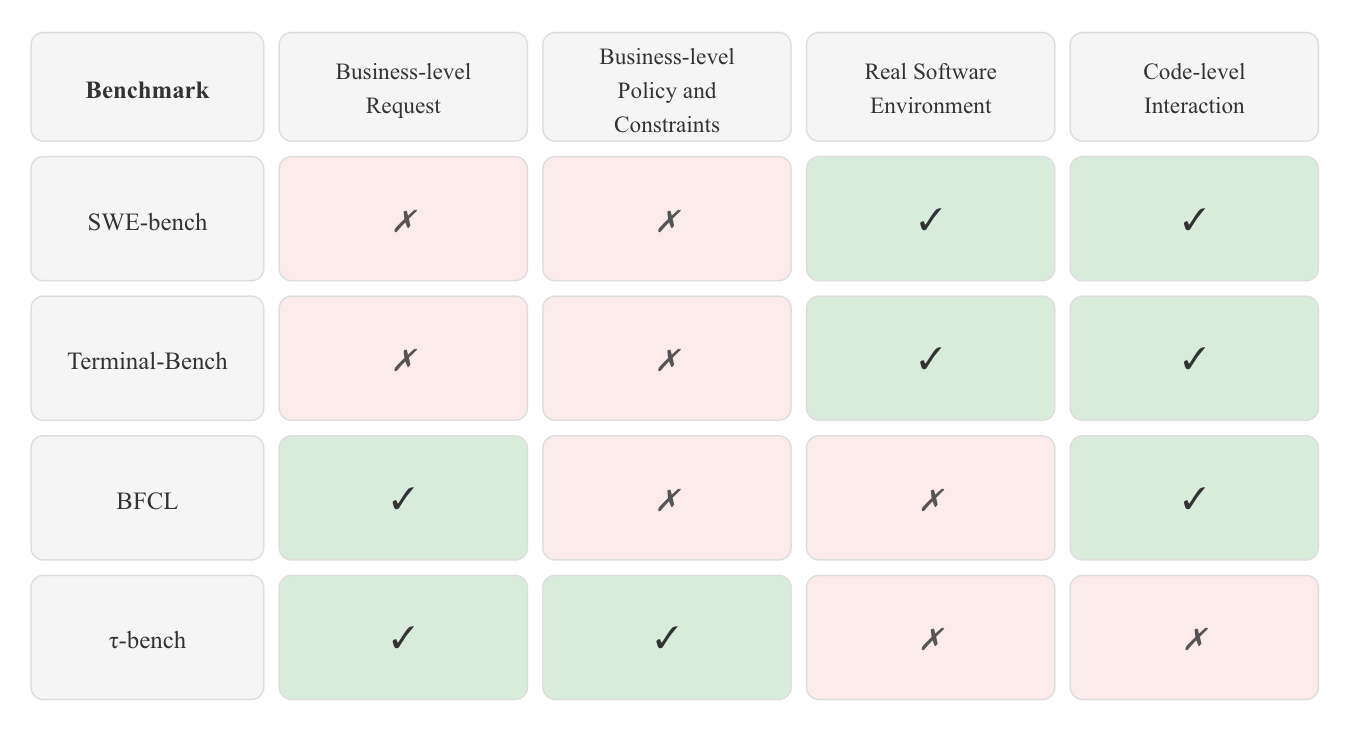}
    \caption{The critical evaluation gap: Code evaluations lack business context; business evaluations lack code interaction.}
    \label{fig:eval-gap}
\end{figure}

Existing evaluations usually stress one or two capabilities at a time: choosing tools, business reasoning, writing code, or manipulating a complex software system. Few require an agent to do everything end-to-end.

\section{Case Study: Coding Agent in Enterprise Resource Planning}

To cover this gap, we ask: \emph{Can a coding agent run business processes end-to-end, under realistic constraints, by writing and executing its own code against a live ERP instance?}

\subsection{Setup}

An ERP is the single source of truth for businesses' operations, supporting more than 3.8 million companies worldwide and, by some estimates, underpinning 77\% of the world's transaction revenue. We identified the ERP as our ideal testing ground for ``general'' business tasks because it contains nearly all core business functions---finance, HR, supply chain, and customer relationship management---as native, interdependent modules. In this environment, we can test how the agent completes simple, single-module tasks like onboarding an employee, but also complex, cross-module tasks like fulfilling a customer order end to end. Furthermore, the ERP has certain validation rules built into its software, while a majority of business logic still remains as the agent's burden.

\begin{figure}[ht]
    \centering
    \includegraphics[width=0.65\textwidth]{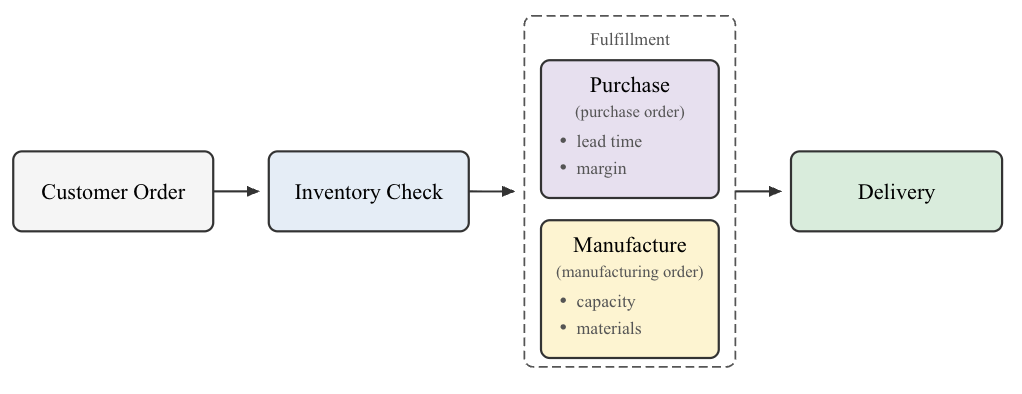}
    \caption{An example of a process in an ERP, involving an order-to-delivery workflow.}
    \label{fig:erp-process}
\end{figure}

\subsection{Environment}

We built a fictional company inside Odoo 19.0 Community Edition, the open-core ERP with over 12 million users across SMBs and large enterprises. Odoo is particularly well suited for this research: unlike most proprietary business apps, it is free to self-host, easy to spin up in many parallel sandboxed instances, and provides visibility into the underlying PostgreSQL database. This gives us a realistic ``world model'' of business operations (sales, inventory, manufacturing, HR, purchasing) while still allowing fine-grained control. To mimic a production environment, we populate the Odoo instance with complete company data---products, vendors, price lists, bills of material, lead times, etc.

We containerize the environment, agent execution, and verification modules separately to prevent information from the verifier from contaminating agent reasoning and to facilitate reproducibility.

\subsection{Task Structure}

We test the agent's ability to complete real-world business workflows that are commonly found in an ERP. Each task gives the agent a natural-language instruction with an objective, constraints, and a policy rulebook. For example:

\begin{lstlisting}[language={}, keywords={}]
Instruction: TechStart Solutions needs 40 ergonomic chairs within 7 days
(budget: $12,000). DesignHub Agency needs 30 chairs within 10 days
(budget: $9,000). Handle both orders, allocating stock and creating
manufacturing orders as needed.

Policy: Maintain at least 25% gross margin. Minimize procurement costs.
You are not allowed to increase the list price of the finished goods.
\end{lstlisting}

In each scenario, the agent has to make at least 2 interdependent business decisions while being subject to at least 2 interacting operational constraints. To pass the example scenario above, the agent must query the ERP, determine the optimal fulfillment strategy, and document its solution (as confirmed sales, manufacturing, and purchase orders)---all through writing and executing code at runtime.

\begin{figure}[ht]
    \centering
    \includegraphics[width=0.5\textwidth]{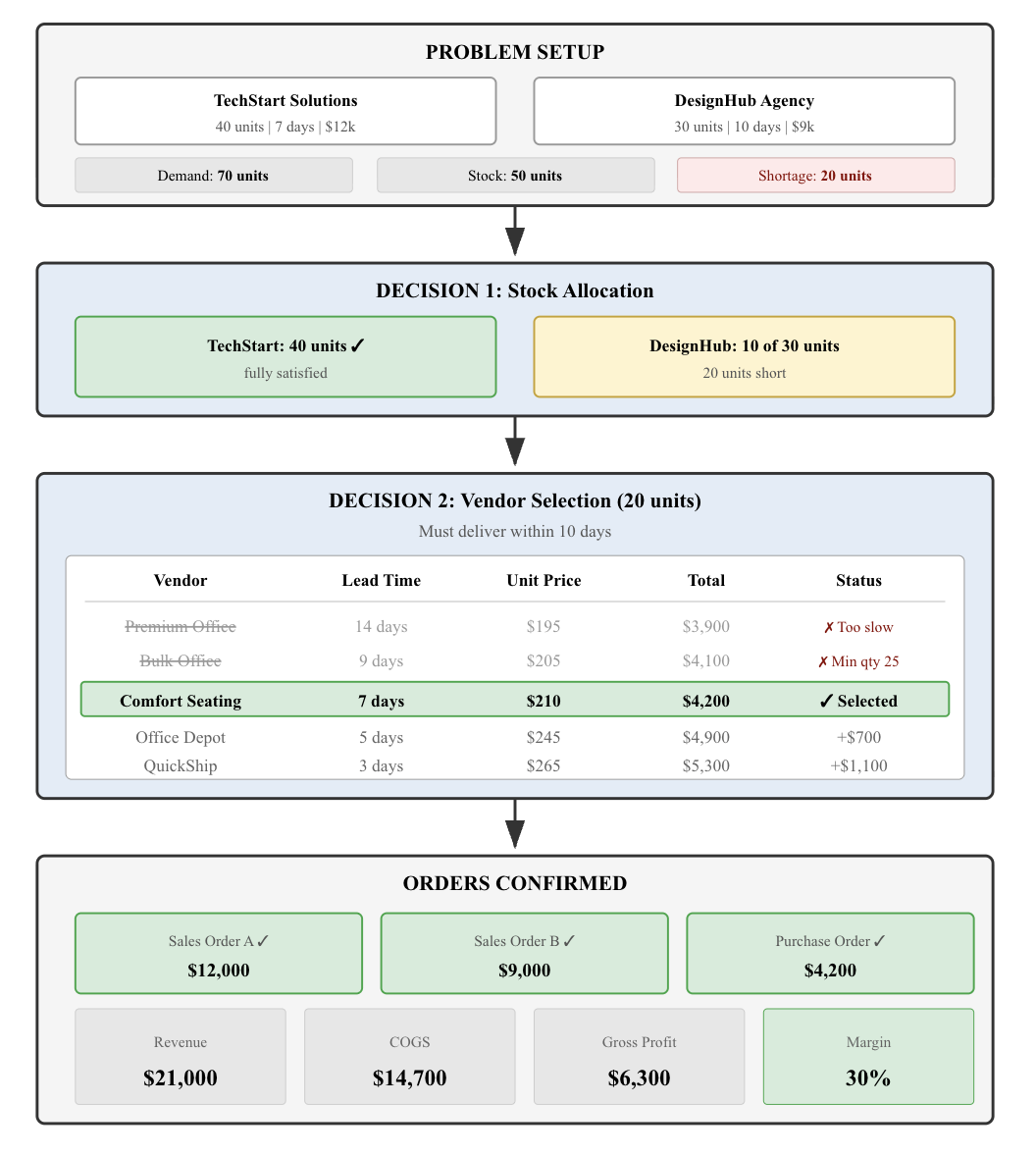}
    \caption{A visualization of the decisions present in a simplified scenario. The solution to this task requires multiple choices and execution steps within the ERP.}
    \label{fig:scenario}
\end{figure}

\subsection{Agent Harness}

We developed a simple coding agent harness for our trials. The harness includes only a bash tool, which is how the agent writes and executes code scripts to interact with the Odoo environment. We provide the agent with database credentials, an isolated workspace for temporary files, and rudimentary examples of five Odoo data models. We do not provide comprehensive documentation on the Odoo instance and API. This compels the agent to iteratively map out the ERP environment from scratch.

We test the agent with GPT-5 and Claude Sonnet 4.5, using maximum allowed reasoning effort/thinking budget settings.

\subsection{Evaluation}

After each trial, our task verifier compares the PostgreSQL database with ground-truth rubrics that define the correct database end-state. Our rubric considers the following:
\begin{enumerate}\itemsep2pt\parskip0pt
    \item \textbf{Constraint Resolution:} Did the agent satisfy all constraints posed by the task instruction and data loaded into the ERP instance?
    \item \textbf{Resource Optimization:} Did the agent solve the task optimally? E.g.\ figured out the most cost-effective fulfillment plan.
    \item \textbf{Traceability:} Are the resulting data objects linked correctly in the ERP? E.g.\ procurement orders are linked to the sales orders they fulfill.
    \item \textbf{Policy Adherence:} Did the agent follow all the rules outlined in the policy rulebook provided to it in the prompt?
\end{enumerate}

\section{Results}

\subsection{Success Out-Of-The-Box}

In the first trial of 10 easy scenarios, our coding agent using Claude Sonnet 4.5 reliably scores above 80\% on the verifiers. These simple tasks include creating sales orders for one or two customers, selecting the cheapest vendor, and generating invoices. This consistent accuracy is impressive: even given limited documentation, the agent intuitively issued correct API calls and navigated Odoo's data model.

The coding agent was so successful on these tasks that, in order to challenge the agent, we needed to scale up the complexity of the tasks to require the agents to make 5+ decisions and weigh 5+ constraints, at which point many domain-level resource allocation tasks became challenging even for humans to work out without computational assistance.

\begin{figure}[H]
    \centering
    \includegraphics[width=0.85\textwidth]{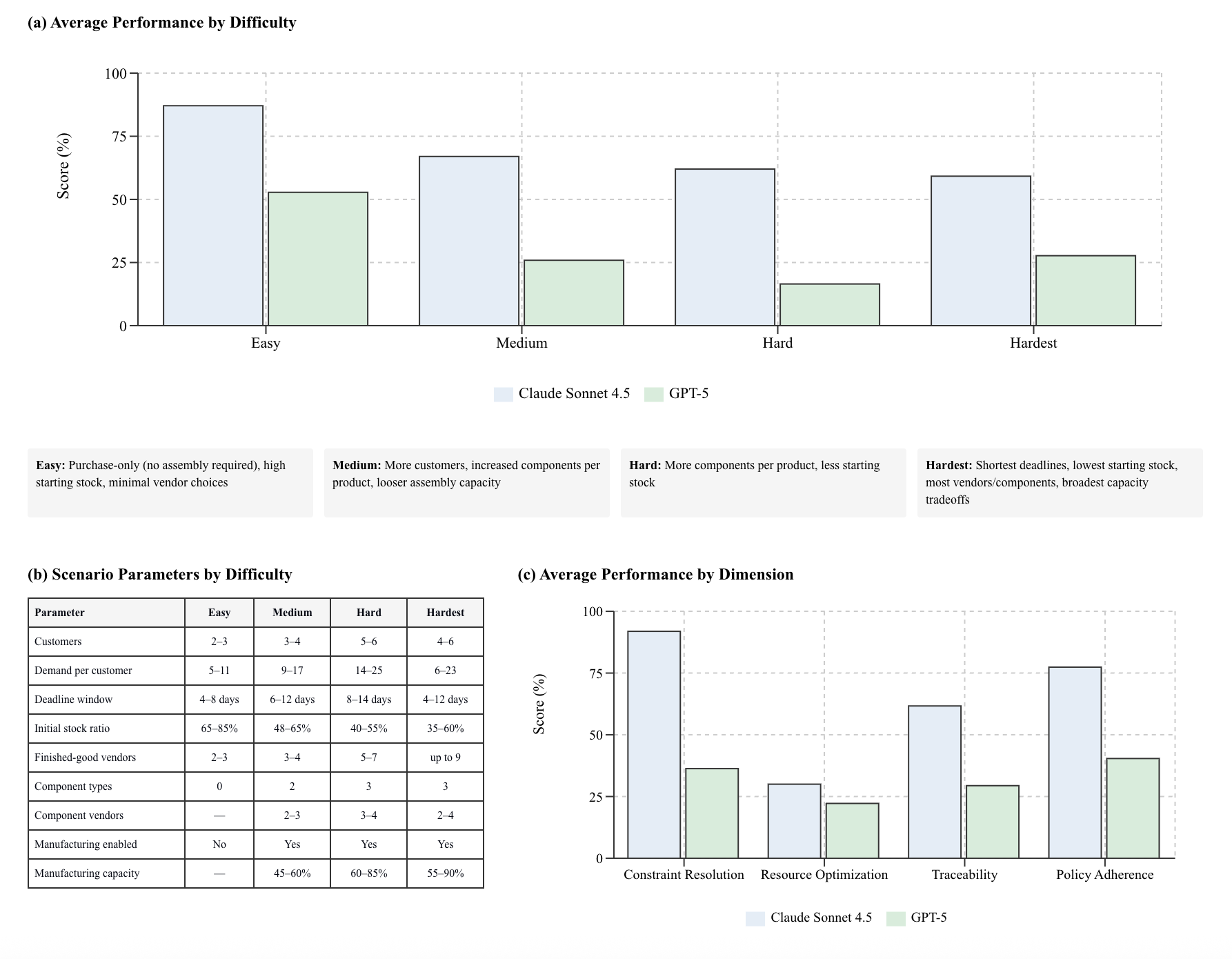}
    \caption{We test Claude Sonnet 4.5 and GPT-5 on 20 scenarios across a gradient of difficulty from Easy to Hardest. We observe that increasing the complexity of the scenarios by increasing the number of constraints results in a breakdown of performance. We separate our evaluation of each scenario into 4 dimensions: Constraint Resolution, Resource Optimization, Traceability, and Policy Adherence. Noticeably, GPT-5 tended to come up with comparable or even better-quality business plans as Claude 4.5 Sonnet, but GPT-5 struggled more with correct API calls, which caused lower scores across the board.}
    \label{fig:results}
\end{figure}

\subsection{Failure Modes}

As complexity increased, characteristic failures began to emerge.

Certain issues came from the business reasoning side alone. Initially, in ``medium'' and some ``hard'' scenarios, the agent produced solutions that were feasible but suboptimal. For instance, all or most constraints were satisfied, but the agent failed to calculate the most cost-effective outcome. Eventually, for the remaining ``hard'' and ``hardest'' scenarios, the agent stops satisfying constraints altogether. The agent's traceability also degrades with additional complexity as it neglects maintaining documentation of its more complex decisions.

Beyond simple reasoning failures, however, we started to see a breakdown in the agent's ability to maintain coherence between the business level and the code level.

\subsubsection{Lazy Code Heuristics}

One class of such problems can be described as ``lazy code heuristics'' that don't accurately implement business logic.

The most salient example involved a task policy to \emph{only order goods and components from American vendors}. Rather than filtering vendors by address, the agent used a glaringly wrong lazy heuristic based on the vendor's name.

\begin{lstlisting}[language=Python]
# Policy in the request:
#   "...order only from American vendors..."
# Relevant Odoo data model:
#   res.partner
#     |-- name
#     |-- contact_address
#     |-- Country Info
#     |     |-- country_code
#     |     +-- country_id
#     +-- Company Info (...)

# Code the agent wrote to filter vendor list
is_american = (
    "American" in vendor.name
    or "Northern" in vendor.name
    or "Catalyst" in vendor.name
    # etc.
)
\end{lstlisting}

The agent correctly understands the natural-language policy (``order only from American vendors'') but implements it with a crude string-matching shortcut. The business intent is correct, but the code that is supposed to enforce it is wrong.

\subsubsection{Hallucinations in the Business Layer}

We also observed the case where the agent hallucinates at the business layer, causing it to write functional code that ultimately leads it astray. In one case, the agent is tasked with scrapping faulty LED Boards that were found to have condensation damage. The agent asserts that if the boards have water damage, they \emph{must} be stored in a fridge, even though the setup actually has them in the warehouse with all others. When it queries the imaginary ``fridge'' location, it finds no results, so it assumes those boards have been discarded.

\begin{figure}[ht]
    \centering
    \includegraphics[width=0.4\textwidth]{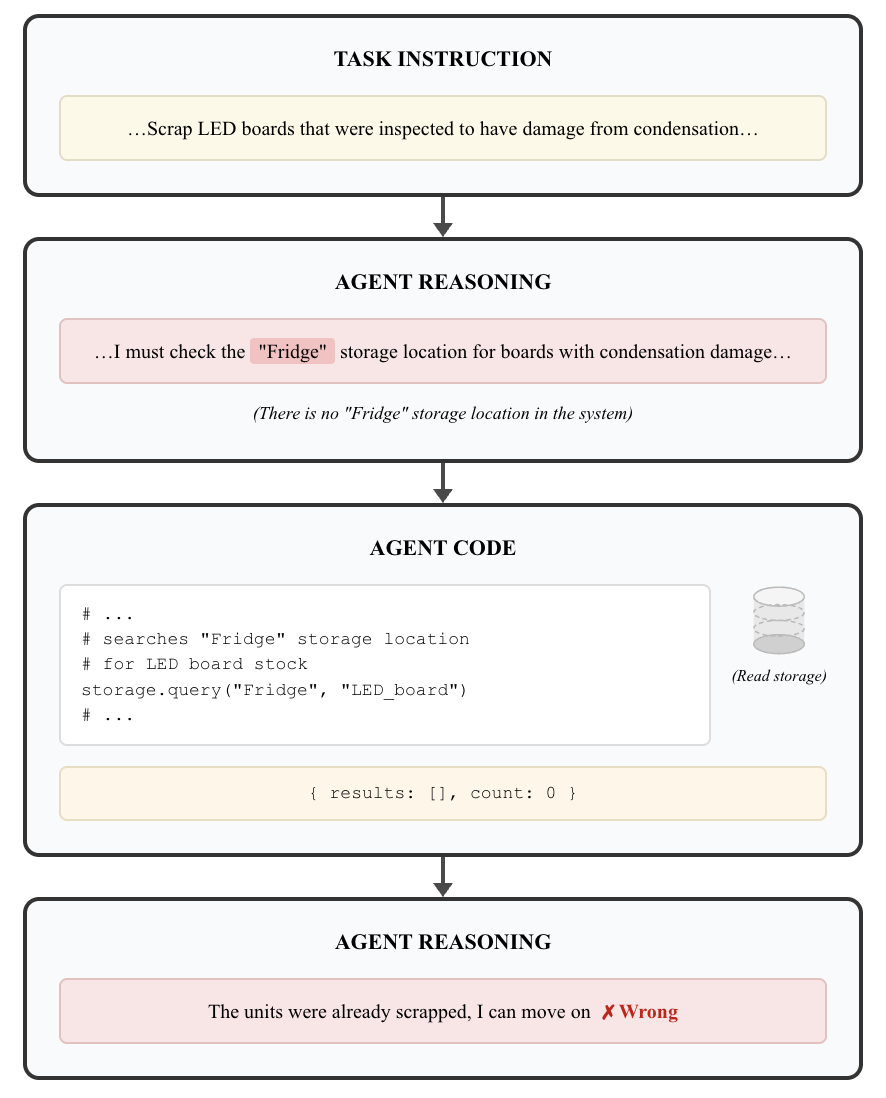}
    \caption{The agent hallucinates that damaged LED boards must be stored in the fridge, without any supporting evidence in the system, leading to functional-but-misleading code.}
    \label{fig:hallucination}
\end{figure}

The direction of failure flips from the previous case. The code is locally coherent given the agent's belief that there is a ``Fridge'' storage location. The query execution makes sense, but the underlying reasoning is hallucinated.

\subsubsection{Ignored Policy Constraints}

In some runs, agents failed to internalize explicit rulebook constraints at all.

For an employee vacation request HR scenario, a policy stating that ``\emph{days off should be consecutive}'' was simply ignored during reasoning, leading to prohibited fragmented schedules.

\begin{figure}[ht]
    \centering
    \includegraphics[width=0.4\textwidth]{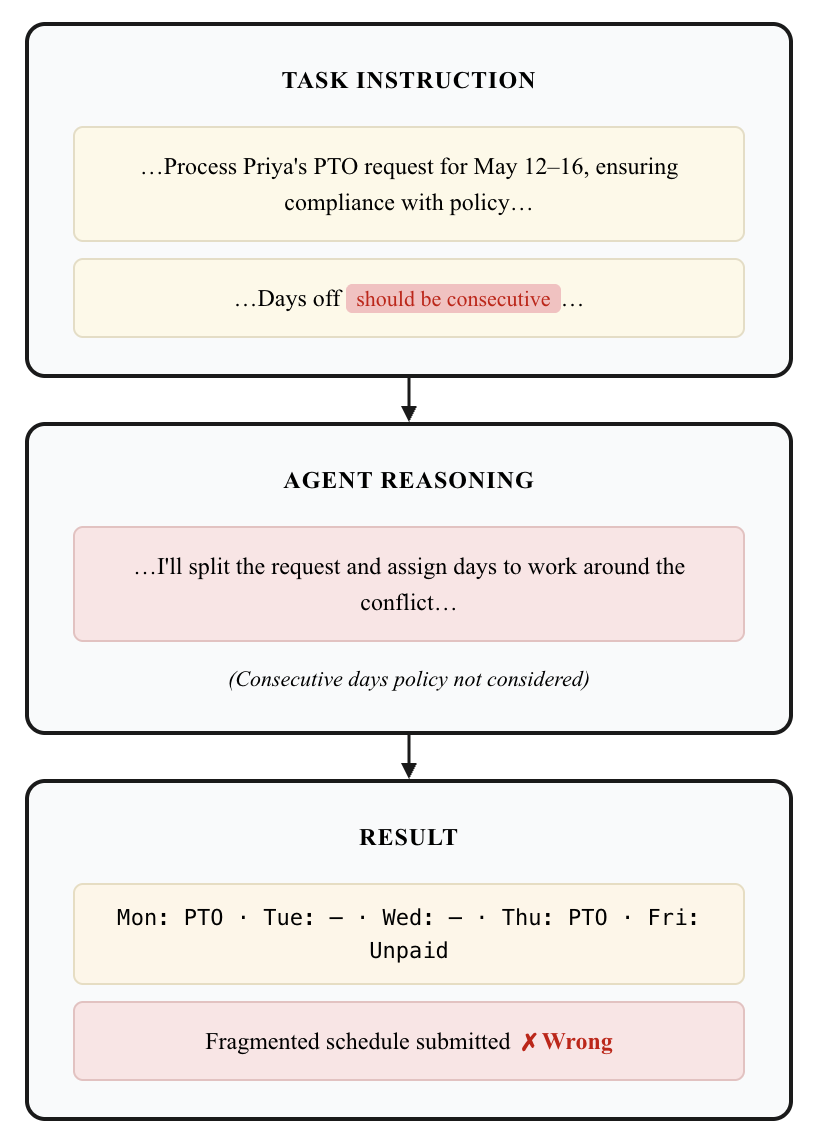}
    \caption{The agent disregards the policy requirement to only schedule consecutive days off, and incorrectly claims successful completion.}
    \label{fig:policy-violation}
\end{figure}

In more complex procurement tasks, margin requirements and lead-time rules were dropped despite being queried correctly from the ERP. Here, the break happens one step earlier: the agent simply does not carry certain rules forward through its reasoning at all.

\subsubsection{Overconfidence}

Whether the solution was optimal, suboptimal, or incorrect, one thing remained constant: the agent almost always reported success and remained unaware of its shortcomings.

This reflects an asymmetry in the feedback profile of the environment. At the code level, the agent receives concrete signals: bad imports, malformed payloads, and incorrect field names, which all produce exceptions. Conversely, we observe that at the business level, feedback is sparse. The inbuilt ERP guardrails reject obviously invalid and ill-formatted interactions but won't flag suboptimal choices or policy violations. Only when we run the task verifier do we see that the outcome was wrong.

This pattern can be viewed as a form of specification gaming: the agent optimizes for the measurable proxy (code execution) rather than the true objective (business correctness). Recent work demonstrates that gaming behaviors generalize, models trained to exploit easily-discovered reward signals will zero-shot transfer those behaviors to novel environments \cite{denison2024sycophancy}. Coding agents may be particularly susceptible: their training provides dense, unambiguous feedback at the code layer (tests pass, scripts execute, errors resolve), effectively teaching that execution success equals task success. This learned prior does not transfer when code is merely the \emph{medium} for a business objective rather than the objective itself. These silent failures point to the untrustworthiness of the agent's own pass/fail conclusions.

\section{Analysis}

The most striking result is that the coding agent can immediately complete straightforward tasks out-of-the-box, at near-human efficacy. With no ERP-specific tooling, the agent reliably executed real business workflows---creating orders, selecting vendors, generating invoices---that would typically require dedicated integrations.

Let us revisit our criteria for a ``successfully'' generalizable coding agent: (1) the agent was able to understand our instructions, (2) inspect the undocumented environment state, (3) reason through the business-level solution, and (4) write a series of scripts to perfectly complete the task and match our verifiers. If all real-world tasks were this simple, we could label the coding agent as ``general'' already.

But when scenarios were loaded with complex decisions and constraints, failures emerged in ways disconnected from raw coding ability. The four failure modes are all different ways of breaking the business-code translation:

\begin{itemize}
    \item \textbf{Lazy heuristics:} correct understanding, incorrect implementation
    \item \textbf{Hallucinations:} correct code, incorrect world model
    \item \textbf{Ignored constraints:} rules dropped before reasoning even begins
    \item \textbf{Overconfidence:} code-level success mistaken for task-level success
\end{itemize}

\textbf{In order to get coding agents to generalize, we should measure and optimize code and business-level correctness together.}

\section{Conclusion}

Coding agents represent a promising path toward general-purpose AI. They are already widely adopted, rapidly improving, and operate in the same medium through which most white-collar work already flows: software.

After examining performance in our simulations, we find that coding agents have advantageous traits that help them succeed in settings where the goal is not to write code, but to achieve correct real-world outcomes \emph{through} code in complex systems. If these agents can learn to reason reliably across unmapped business domains while retaining their native fluency in code, the distance to broad automation shrinks considerably.

Today, coding agents still do not generalize to complex business workflows. Domain-specific tools and guardrails will remain important in the near term---but they address failure modes one instance at a time. This risks running afoul of the bitter lesson \cite{sutton2019bitterlesson}, and history suggests that general methods eventually win.

\bibliographystyle{unsrt}
\bibliography{assets/bibliography/2026-04-27-coding-agents}

\end{document}